\documentclass[12pt]{article}
\usepackage{amsmath}
\usepackage{euscript}
\usepackage{latexsym,graphics,epsfig}
\usepackage{latexsym}
\usepackage[all, knot]{xy}
\usepackage{amsfonts,amssymb}
\usepackage[all]{xy}
\title{Topos Mediated Gravity: Toward the Categorical Resolution of the Cosmological Constant Problem}
\author{Jerzy Kr\`ol $ ^{\ast}$ \\  Institute of Physics, 
University of Silesia, \\ Uniwersytecka 4, Pl 40007 Katowice,\\  Poland}

\begin{document}
\baselineskip5mm
\maketitle
\begin{abstract}
\baselineskip4mm
According to D\"oring and Isham the spectral topos corresponds to any quantum system. The description of a system in the topos becomes similar to this given by classical theory, up to multiplication of observables. Logic of the emergent theory is rather intuitionistic than classical. Adding to the language the type of self-adjoint operators and their products, which are interpreted by the on-stages daseinisations and their products, one gets commuting presheaves in the topos. The interpretation agrees with on-stages daseinisations of squares up to some automorphisms of the topos. The uncertainty principle does not hold for such an interpretation. According to the recent proposition by the author, topoi can modify local smooth spacetime structure. A way how to add gravity into the spectral topos of a system is presented. Assuming that a quantum system modifies the local spacetime structure and interacts with a gravitational field via the spectral topos, the lowest energy modes of a quantum harmonic oscillator are gravitationally nullified. Moreover, a theory of gravity and systems should be symmetric with respect to the 2-group of automorphisms of the category of topoi. Under the $\delta_0$-interpretation for the hamiltonian of a quantum harmonic oscillator, its lowest modes have vanishing contributions to the cosmological constant. This is related with a fundamental higher symmetry group of gravity. Without the $\delta_0$-interpretation, small non-zero value of the effective cosmological constant can be approached.

\end{abstract}
%PACS number(s): 

\vspace{10mm}

$ ^{\ast}$ e-mail: iriking@wp.pl

\newtheorem{corollary}{Corollary}
\newtheorem{definition}{Definition}
\newtheorem{lemma}{Lemma}
\newtheorem{theorem}{Theorem}
\newpage
\section{Introduction}
The cosmological constant problem (cc problem) is one of the major mysteries in present-days physics (see e.g. \cite{weinberg1989,straumann2002,padmanabhan2008}). This is also one of the most often discussed problems of physics and the attempts to resolve it are numerous. In fact, one should speak about some independent ingredients of what is called cc problem. Roughly, the cc problem consists in the huge contribution of the lowest point energy of quantum systems and fields to the cosmological constant (cc) and the perfect cancellation between the Einstein's cc term and density of matter and energy, which occurs exactly at our epoch. Second ingredient of the problem is the question why this cancellation is not exact but rather cc has a very small positive value.

The effective cosmological constant can be expressed as $\Lambda = 8\pi G\rho_{vac}+ \Lambda_0$ where $\Lambda_0$ is the cosmological constant appearing in Einstein's equations and $\rho_{vac}$ refers to the vacuum energy density determined by quantum systems \cite{straumann2002}. There exist experimental limitations on the upper limit of the value $\Lambda$ and the precise cancellation between $\Lambda_0$ and $\rho_{vac}$ is the true challenge and is one of the ingredients of the cc problem. Moreover, even though one cannot calculate the vacuum energy density in QFT, the estimation of the order of the energy which would be valid gravitationally exceeds in many orders this what can be expected. This is also the part of the cc problem. 

It is often stated that a new kind of symmetry should appear in order to cancel gravitational contributions to the cosmological constant, eventually below the supersymmetry breaking scale. 

In this paper we address two issues connected with the cc problem. These are the vanishing contributions of the lowest-points energies of quantum systems in spacetime to the cosmological constant, and appearance of the new symmetry of gravity. The technique we employ here to both of the issues is rather unusual and is based on recent work on topoi by D\"oring and Isham \cite{isham2007a,isham2007b,isham2007c,isham2007d} and the present author \cite{krol2006a,krol2006,krol2007}. Some motivations derive from work on higher categories and Yang-Mills theories and their role in quantum and general relativity and physics in general \cite{crane2006,baez2004,cattaneo2002}. This parallels the program of the categorification in pure mathematics (see e.g. \cite{baez1998}).     

The role of the spectral topos in the paper is different to that in D\"oring and Isham approach, in the sense that we do not assume the existence of a complete interpretation of a quantum theory in the topos as a (semi)-classical theory. Rather, we consider the spectral topos as some category which deforms both a quantum theory and the local spacetime structure. The deformation of a quantum theory, however, does not give rise to a complete semi-classical theory. Moreover, the on-stages reasoning in the Grothendieck topos, is considered as giving the approximations for various, to be interpreted, external results. The deformation of quantum theory follows the internal interpretation of the formal language in the topos where the symbols $P$ and $X$, corresponding to the momentum and the position observables, are represented by the morphisms $\breve{\delta}(A):\underline{\Sigma}\to R_{\cal{H}}$. However, we do not assume the consistent interpretation of the others self-adjoint operators like $P^2$ or $X^2$ as the daseinisations-morphisms. Also, the non-self-adjoint products like $PX$ or $XP$ may not be consistently, with the daseinisations, represented in the topos. Besides, there are no symbols for the products in the language in the original D\"oring-Isham approach.   

The difficulty with adding consistently the daseinisations of all self-adjoint operators of a quantum theory to the interpretation of the theory relays on the fact that the preasheaf $R_{\cal{H}}$, corresponding to the real numbers in D\"oring and Isham approach, is not a commutative ring. Hence, the multiplication of $\breve{\delta}(A)\breve{\delta}(B)$ and subtraction $\breve{\delta}(A)-\breve{\delta}(B)$ are not, in general, well-defined. The interpretations of $AB$ do not agree with those of $A$ and $B$ given by $\breve{\delta}(A)$ and $\breve{\delta}(B)$. This would mean that we do not even have well defined candidates for the interpretations of expressions like $AB$ and $A-B$. Even in the case of $AA$, where $\breve{\delta}(AA)$ is defined, one has not $\breve{\delta}(A)\breve{\delta}(A)$ well defined. This is particularly harmful when one wants to study commutation relations of observables in the topos. If the spectral topos could interpret whole quantum theory such that all observables are represented by morphisms with values in a commutative ring and such that this would agree with the interpretations of products and subtractions of operators, then a quantum theory would become internally classical (not quantum) and intuitionistic. In such a representation, every pair of observables represented in the spectral topos would commute. However, this is not the case. The preasheaf corresponding to the object of real numbers is merely an additive group.
 
That is why in the paper we do not make use of the morphisms $\breve{\delta}$'s nor we assume the existence of the complete interpretation of all quantum observables, by these. This more that in the formal language, which is assigned to a quantum system and interpreted in the topos, one does not have the symbols corresponding to $A^2$ or $AB$. Instead, we attach the symbols $AB$ or $A^2$ to the language as having the type of operators, and interpret them as on-stages defined operators\footnote{$\delta_V(A)$ is the (outer) daseinisation of a self-adjoint operator $A$ on stage $V\in {\cal{V}}$ (see Sec. \ref{secIE})} $\delta_V(A)\delta_V(B)$ and $\delta_V(A)\delta_V(A)$, $V\in {\cal{V}}$. This does not agree with the morphisms $\breve{\delta}$'s. Even though the morphism $\breve{\delta}(A^2)$ cannot be built as consistent with others $\breve{\delta}$'s, one still can consider the on-stages representations for $AB$ and $A^2$ as $\delta_V(A)\delta_V(B)$ and $\delta_V(A)\delta_V(A)$ respectively. This interpretation does not assign the morphisms $\breve{\delta}$'s to the observables which would extend on multiplication of observables. It interprets the symbols like $AB$ or $A^2$ of the type of operators, by on-stages defined operators. Thus, we do not claim that all self-adjoint operators are consistently interpreted by morphisms-daseinisations. Rather, the spectra of these interpreted operators have physical meaning for gravitational interactions with a quantum system interpreted in the spectral topos. However, the emerging symmetry of this interactions should be suitably, categorically extended. The on-stages interpretation, as described above, we call $\it pre-semi-classical$ in the paper. This is not the semi-classical interpretation in D-I sense. 

In Sec. \ref{secIE} I present the formulation of the theory of a quantum oscillator internally in the spectral topos $SET^{{\cal{V}}^{op}}$ of presheaves on the category ${\cal{V}}$ of unital boolean algebras of operators on the oscillator's Hilbert space ${\cal{H}}$ \cite{isham2007a}. This follows the construction by D\"oring and Isham of a formal language assigned to any quantum system and its spectral topos \cite{isham2007a,isham2007b,isham2007c}, however the interpretation I use is the pre-semi-classical as above. Under this interpretation quantum operators commute on-stages, hence these commute in the topos. The theory of a quantum oscillator becomes similar to the semi-classical theory as in D\"oring and Isham, up to the point where the interpretation of products of self-adjoint operators is considered.

I also analyse the possibility that although the difference between $\breve{\delta}(A)\breve{\delta}(A)$ when defined on stages, and $\breve{\delta}(A^2)$, cannot vanish in general, this can be gauge out by some higher categorical symmetry.         
   
Observables of a quantum oscillator in $SET^{{\cal{V}}^{op}}$ correspond to commuting operators on every stage $V\in {\cal{V}}$, hence they commute in $SET^{{\cal{V}}^{op}}$. However, they do not give rise in general to the corresponding consistent ${\breve{\delta}}$'s morphisms. The internal energy spectrum of the observables has lowest-points value of energy equal to $0$ at every stage $V$, such that at this stage the daseinisations of the operators $X$ and $P$ have $0$ in their spectra. If the other stages of the hamiltonian are set to zero, the constant zero presheaf $\tilde{0}$ becomes the lowest value of the internal energy spectrum of the harmonic oscillator.

The internalisation in some other boolean topoi allows to describe categorical gauge symmetries such that the commutation relation between $X$ and $P$ holds only in the weak form, i.e. up to a gauge. Hence, the uncertainty principle holds only up to a gauge. This shows the relevance of categorical gauge symmetries in the topos approach to the cosmological constant problem.     

In Sec. \ref{secGR} gravity is added to the topos of a quantum system. A quantum system is considered in spacetime and the spectral topos of the system modifies locally spacetime structure according to the work by the author \cite{krol2006a,krol2006}. When gravity is described internally in $SET^{{\cal{V}}^{op}}$, the interactions of the system with classical gravitational field take place internally in the topos. The lowest points of energy have vanishing contributions to cc. 

Sec. \ref{2GR} contains sketch of the formalism applied to any family of quantum systems in spacetime. Some 2-group of symmetries for gravity emerges. 

A brief discussion of the issue of small non-zero positive value of cc in the formalism of topoi and modified spacetimes, appears also in Secs. \ref{secGR} and \ref{2GR}. Detail treatment is postponed to a separate paper.

\section{\label{secIE} Internal Energy Spectrum of a Quantum Harmonic Oscillator}
According to D{\"o}ering and Isham \cite{isham2007a} one can associate a formal language to any physical system. To get an appropriate theory describing the system the language should be interpreted in some categories. The language in question is the typed higher order one. The category is some topos. In the case of the quantum theory of a system, the topos is the spectral topos of presheaves $SET^{{\cal{V}}^{op}}$ on the category ${\cal{V}}$ of abelian, unital algebras $V$ of operators on a Hilbert space ${\cal{H}}$ of the system \cite{isham2000a,isham2007b}. Thus, the language is the local language assigned to the topos (see Bell \cite{bell1988}, Lambek and Scott \cite{lambek1994}).
This higher order typed language ${\cal{L_{H}}}$ corresponding in particular to a harmonic oscillator, is described in the Appendix.

The formal language ${\cal{L_H}}$ of a quantum oscillator $({\cal{H}}, H=P^2/2m+(1/2)m\omega^2X^2, P, X)$, contains the functional symbols $F_{P},F_X,F_H:\Sigma\rightarrow {\cal{R}}$, corresponding to the observables $X$, $P$, $H$ and $m$, $\omega$ are symbols for constant numerical parameters. $P$, $X$ are usual momentum and position self-adjoint operators acting on ${\cal{H}}$.
$\Sigma$ is a symbol for the space of states, classical or quantum \cite{isham2007a}.  

A highly non-trivial fact was established by D{\"o}ring and Isham that the language of a quantum system can have quantum observables represented by the functional symbols which formally looks like the classical case. This is possible since the language has a model or representation in the spectral topos where this is precisely realized \cite{isham2007a,isham2007b,isham2007c}. However, to build such a model one has to allow for the intuitionistic logic of the theory. 

The language of the intuitionistic higher order typed theory should be interpreted in the topos $SET^{{\cal{V}}^{op}}$ derived from the Hilbert space ${\cal{H}}$ of the system. The interpretation was described in \cite{isham2007a,isham2007b,isham2007c} and some elements of this can be found in the Appendix.

The functional symbols, when in the topos, correspond to its arrows (morphisms) between some presheaves $\overline{F}_P,\overline{F}_X,\overline{F}_H:\underline{\Sigma}\rightarrow R_{\cal{H}}$ which are obtained from the self-adjoint operators via the special procedure, called by D{\"o}ring and Isham, daseinisation \cite{isham2007c}. This kind of daseinisation, which leads to morphisms from observables, is denoted by $\breve{\delta}$.  $\underline{\Sigma}$ is the spectral presheaf in the topos $SET^{{\cal{V}}^{op}}$ which interprets the symbol $\Sigma$ of the language \cite{isham2007b}. $m, \omega$ are interpreted by constant presheaves i.e. the real numbers in the topos $\tilde{m}, \tilde{\omega}$. $R_{\cal{H}}$ is the special presheaf called a quantity-value object \cite{isham2007c}. Thus, $\breve{\delta}(A): \underline{\Sigma}\to R_{\cal{H}}$ where $A$ is a self-adjoint operator. The important thing here is that $R_{\cal{H}}$ is not the real number object in the topos. This is rather a carefully constructed presheaf. There is also the other kind of daseinisation which is important at some stages of the constructions, namely {\it outer} and {\it inner} daseinisations, $\delta^{out}$, $\delta^{inn}$ \cite{isham2007c}. These are presheaves assigned to observables which are on-stages represented by self-adjoint operators from some unital boolean subalgebras $V$ of operators. These algebras are the stages of the spectral topos. In the paper we usually refer to the outer daseinisation and omit the index {\it out}. Hence, the outer daseinisation of a self-adjoint operator $A$ at stage $V$ will be denoted as $\delta_V(A)$.  

In a classical theory, the subobjects of space of states correspond to the propositions about a system. In a quantum theory these form, in general, the lattice of projections on the closed subspaces of the Hilbert space of a system. This lattice is non-distributive. If the lattice were a distributive algebra, the observables of the system would commute (see Lemma \ref{lemma1} below). As the consequence, the uncertainty principle would fail to hold and, in this sense, the theory would become rather classical.

In the case of our theory of quantum harmonic oscillator interpreted in the topos $SET^{{\cal{V}}^{op}}$, there exists similar to the classical case, correspondence: propositions about the system correspond to the subobjects of the object of states $\underline{\Sigma}$. This is, among others, the feature which allows for calling such a quantum theory when interpreted in $SET^{{\cal{V}}^{op}}$, semi-classical or neo-realistic one. The subobjects, however, form a Heyting algebra rather than boolean one and the truth value object is the sheaf $\underline{\Omega}$ - subobject classifier of the topos - and appears as the interpretation of the language type symbol $\Omega$.
 
Let us discuss now the difficulty with a semi-classical description of a quantum harmonic oscillator such that this would lead to commuting $X$, $P$ in $SET^{{\cal{V}}^{op}}$.
\begin{lemma}\label{lemma1}
There exists a pair of non-commuting self-adjoint operators on a Hilbert space ${\cal{H}}$ iff the lattice of projections on ${\cal{H}}$ is non-distributive.
\end{lemma}
If there are 2 non-commuting self-adjoint operators there has to exist a pair of non-commuting projections from their spectral families, and then the non-distributivity follows, since non-distributivity is, in fact, equivalent to the existence of two non-commuting projections \cite{degroote2005}. Conversely, if the lattice of projections is non-distributive then two non-commuting projections have to exist and there are two non-commuting self-adjoint operators. $\square$

However, to conclude that in $SET^{{\cal{V}}^{op}}$ every pair of self adjoint operators commute one needs an internal Hilbert space and internal projections. In DI approach this is difficult to achieve, since the object of real numbers, according to which one defines quantum observables in a semiclassical way, is not a commutative ring, hence one cannot build the Hilbert space based on these reals in the topos.

From the other side on every stage $V\in {\cal{V}}$, $(X)_V(P)_V=(P)_V(X)_V$ where $(X)_V=\delta_V(X)$, $(P)_V=\delta_V(P)$ and $\delta_V{A}$ is the (outer) daseinisation of the self-adjoint operator $A$ with respect to $V$. $(B)$ is the internal in the topos version of a not necessary self-adjoint operator $B$.  
However, the commutation of $X$ and $P$ on stages in $SET^{{\cal{V}}^{op}}$ does not mean that $XP=PX$ holds true in the topos. The reason is that $PX$ and $XP$ internalized in $SET^{{\cal{V}}^{op}}$, i.e. $(PX)$ and $(XP)$, cannot be on stages given by $(P)_V(X)_V$ and $(X)_V(P)_V$, as above, due to the not preserving the algebraic structure by the daseinisation operation. Thus, $(PX)$ and $(XP)$ in $SET^{{\cal{V}}^{op}}$ are not related with the operator defined on stages by $(P)_V(X)_V$ and $(X)_V(P)_V$. Moreover, $(P)_V(X)_V=(X)_V(P)_V$, $V\in {\cal{V}}$ while in general $(PX)_V\neq (XP)_V$. Besides, a daseinisation of $PX$ and $XP$ is not defined, since these multiplications are not self-adjoint operators. 

One is faced with the difficulty to define consistently internal in $SET^{{\cal{V}}^{op}}$ $PX$, $XP$ and multiplications like these which would agree with the daseinisation of self-adjoint operators and respect somehow the multiplicative structure.

There exists the possibility that there are no consistent interpretations of $PX$ and $XP$ in the above sense. In that case one has a problem with interpreting the whole content of a quantum theory such that all self-adjoint operators are interpreted as morphisms $\breve{\delta}:\underline{\Sigma}\to R_{\cal{H}}$ and this would agree with interpretations of their products and subtractions. Even so, one can always take the internal representatives of $P^2$, $X^2$ and the like in $SET^{{\cal{V}}^{op}}$ as interpretations of self-adjoint operators, which are again self-adjoint operators and which do not necessarily have corresponding representations by the above semi-classical arrows. These operators can be defined on-stages as $\delta_V(X)\delta_V(X)$ and $\delta_V(P)\delta_V(P)$ correspondingly. Similarly, the interpretations of $PX$ and $XP$ can be given on stages as $(XP)_V=\delta_V(X)\delta_V(P)$ and $(PX)_V=\delta_V(P)\delta_V(X)$ respectively. This allows for the extending of interpretations consistent with the daseinisations of self-adjoint operators on products of operators, however this is not any consistent extension of the interpretation of self-adjoint operators by the daseinisations-arrows in $SET^{{\cal{V}}^{op}}$. Thus, the truncation of the D-I semiclassical interpretation is used. The self-adjoint operators $P$, $P^2$, $X$, $X^2$ etc., are represented as presheaves defined on stages by outer daseinisations or the squares of these. These are not further interpreted by semi-classical daseinisations - morphisms. These last can correspond eventually to self-adjoint operators but we do not make use of the correspondence nor its consistency.  

This interpretation of $X$, $P$, $XP$ and $PX$ I call {\it pre-semi-classical} or {\it $\delta$-interpretation} of quantum operators in $SET^{{\cal{V}}^{op}}$ and is denoted by $(X)_{\delta}$, $(P)_{\delta}$, $(PX)_{\delta}$ and $(XP)_{\delta}$ respectively. Note that $((X)_{\delta})_V=\delta_V(X)$, $((P)_{\delta})_V=\delta_V(P)$, however $(PX)_{\delta}\neq \delta(PX)$, since $\delta(PX)$ is not well defined. In what follows, I will not, in fact, make any essential use of the semi-classical interpretation of observables by arrows in $SET^{{\cal{V}}^{op}}$.     

When defining on stages $PX$ and $XP$, the resulting presheaves do not necessarily correspond to $(PX)$ and $(XP)$ from other possible interpretations. In particular one may not be in agreement with the representation of the observables like $\breve{\delta}(A^2)$ by the functor $\underline{\Sigma}\to R_{\cal{H}}$. 

Even though $(PX)\neq (XP)$ in general, and even in the case of the lack of $(PX)$, $(XP)$ consistent with the daseinisations of $P$ and $X$, the corresponding preasheaves, $(PX)_{\delta}$ and $(XP)_{\delta}$, fulfil   
\begin{lemma}\label{lemma2}
$(PX)_{\delta}=(XP)_{\delta}$ in $SET^{{\cal{V}}^{op}}$.
\end{lemma}
$((PX)_{\delta})_V=\delta_V(P)\delta_V(X)=\delta_V(X)\delta_V(P)=((XP)_{\delta})_V$, $V\in {\cal{V}}$, hence the result follows. $\square$
  
It is obvious that Lemma \ref{lemma2} holds for any pair $A$, $B$ of quantum operators interpreted in the topos $SET^{{\cal{V}}^{op}}$ and whoose multiplication is defined on stages by the corresponding multiplication of daseinized operators. Thus, for the observables of a quantum system with their $\delta$-interpretation in the topos $SET^{{\cal{V}}^{op}}$, it holds
\begin{corollary}\label{cor1}
Under $\delta$-interpretation of observables in the topos $SET^{{\cal{V}}^{op}}$, the uncertainty principle does not hold.
\end{corollary}
However, the interpretation of the product of self-adjoint operators $AB$ and $BA$ is enforced to be $(AB)_{\delta}$ and $(BA)_{\delta}$ rather than some other $(AB)$ and $(BA)$. For the later, $(AB)\neq (BA)$ holds true usually.  

We can now approach the energy spectrum of the harmonic oscillator in $SET^{{\cal{V}}^{op}}$. We do this at every stage $V\in {\cal{V}}$ not assuming any longer the validity of the internal canonical commutation relations (see Corollary \ref{cor1}). However, we do assume the $\delta$-interpretation for products of operators. 

In terms of daseinisation operation $\delta_V$, given two (possibly, non-commuting) self-adjoint operators $A$, $B$, while taking their (outer) daseinisations on-stages one can use these to generate spectra of, say, $AB$, $A^2$ or the like, as the spectra $\delta_V(A)\delta_V(B))$, $\delta_V(A)\delta_V(A)$, correspondingly, on each stage $V\in {\cal{V}}$.
\begin{definition}
The energy spectrum of a quantum system in $SET^{{\cal{V}}^{op}}$ obtained as above is the on-stages {\rm internal energy spectrum} of the system.
\end{definition}
Note that the internal spectrum is not, in general, any presheaf, hence it is neither an object of the spectral topos, nor any interpretation of the external spectrum of the system.
\begin{lemma}\label{1,3}
If $X$ and $P$ have $0$ in their spectra, on some stage $V\in {\cal{V}}$, then the Hamiltonian $H$ of the quantum system has $0$ in its (on-stages) internal energy spectrum on $V$. 
\end{lemma}
This follows from the interpretation on-stages of the Hamiltonian $H$ by commuting operators $\delta_V(P)$ and $\delta_V(X)$. $\square$

For classical systems, this is certainly not true in $SET$. 

Let us set as zero those $H_V$ where $P_V$ and $X_V$ do not have zero in their spectra. The corresponding interpretation will be called $\delta_0$-interpretation of $H$. We will return to the discussion of this interpretation in Sec. \ref{2GR}. 

For the $\delta_0$-interpretation $H$, the zero-point value of energy modes are distinguished.
\begin{lemma}{\label{lemma3}}
Supposing $\delta_0$ interpretation, the lowest value of the internal energy spectrum of a harmonic oscillator $({\cal{H}}, H=P^2/2m+(1/2)m\omega^2X^2, P, X)$ in $SET^{{\cal{V}}^{op}}$ is the constant presheaf $\tilde{0}$. 
\end{lemma}
At every stage $V\in {\cal{V}}$, $H_V=\delta_{0,V}(P)\delta_{0,V}(P)+(1/2)m\omega^2\delta_{0,V}(X)\delta_{0,V}(X)$ and the spectrum of $H_V$, $\sigma_V(H)$, contains the value $0$ since $\delta_{0,V}(P)$ and $\delta_{0,V}(X)$ commute and $0\in \sigma_V(P)$ and $0\in \sigma_V(X)$. If $0\notin \sigma_V(P)$ or $0\notin \sigma_V(X)$ the $\delta_0$-interpretation gives $0\in \sigma_V(H)$. Hence, at every stage $V\in {\cal{V}}$ the lowest point of the energy spectrum is $0$. This defines the constant preasheaf $\tilde{0}$. $\square$

Thus, $P$ and $X$ represented in $SET^{{\cal{V}}^{op}}$ as the presheaves $\delta_V(P)$ and $\delta_V(X)$, do commute in $SET^{{\cal{V}}^{op}}$. Certainly, $PX$ and $XP$ represented in $SET^{{\cal{V}}^{op}}$ in another way, need not be equal. Hence we lost the connection of the external $PX$ and $XP$ with the internal $(PX)$ and $(XP)$. We, however, preserve the external connection of $PX$, $XP$ with $X$ and $P$ also internally on stages. 

Let us approach somehow the gap between the interpretations $(PX)$ and $(XP)$ of true $PX$ and $XP$ and on-stages daseinisations $\delta_V(P)$ and $\delta_V(X)$, $V\in {\cal{V}}$.
In general $(PX)_V\neq (XP)_V$, $V\in {\cal{V}}$, however, let us suppose that 

(S1): $(PX)_V\in V$, $(XP)_V\in V$, $V\in {\cal{V}}$.

(S2): There exist automorphisms $f_V$ of each $V\in {\cal{V}}$, such that $f_V{|V_{sa}}$ is an automorphism of $V_{sa}$ and $f_V((PX)_V)=(XP)_V$, where $V_{sa}$'s are the subalgebras of the self-adjoint operators at every stage $V$. 

The automorphisms $f_V$ of every algebra $V_{sa}\subset V\in {\cal{V}}$ define automorphisms of every presheaf in $SET^{{\cal{V}}^{op}}$. Thus, we get an automorphism $F$ of the topos $SET^{{\cal{V}}^{op}}$. 
Let us consider $f_V(\delta_V(X))$ and $f_V(\delta_V(P))$. Obviously, $f_V(\delta_V(X))f_V(\delta_V(P))=f_V(\delta_V(P))f_V(\delta_V(X))$. Moreover, $f_V((XY)_V)=f_V((YX)_V)$ on every stage $V$. Now, define the interpretation of a quantum theory (quantum oscillator) such that the modified daseinisation (interpretation) of $X$ at the stage $V\in {\cal{V}}$ will be $f_V((\delta_V(X)))$. Same for $P$. Let us call this interpretation the $(\delta,F)$-interpretation. This makes that $X$, $P$ commute on-stages and this agrees with the commutation of the interpretations of $XP$ and $PX$.
Thus, under the suppositions (S1) and (S2) one has
\begin{lemma}\label{1,5}
$(PX)_{(\delta,F)}=(XP)_{(\delta,F)}$ in $F(SET^{{\cal{V}}^{op}})$ for some automorphism $F$ of $SET^{{\cal{V}}^{op}}$. 
\end{lemma}
Is there any possible physical meaning given to the internal weak equality, $(PX)\approx (XP)$, which holds up to some automorphism $F$ of $SET^{{\cal{V}}^{op}}$? The following reformulation of the Lemma \ref{1,5} helps answering   
\begin{corollary}\label{cor1}
Under the suppositions (S1) and (S2), the uncertainty principle in $SET^{{\cal{V}}^{op}}$ holds in the weak form, i.e. up to some automorphism $F$ of the topos.
\end{corollary}
The advantage of dealing with $F(SET^{{\cal{V}}^{op}})$ is that one can have the interpretation of a quantum theory where $X$ and $P$ commute, hence the uncertainty principle may not be true. $X$, $P$ are now interpreted by $f_V(\delta_V(X))$ and $f_V(\delta_V(X))$.

Moreover, the definition of the modified daseinisation of non-self-adjoint products of operators, like $PX$, can be determined on stages as $f_V(\delta(P)_V)f_V(\delta(X)_V)$ up to some automorphism $F$ of $SET^{{\cal{V}}^{op}}$. 
We obtained the commutativity in the topos up to some automorphism of the topos. Again, the $\delta_0$-interpretation for $H$ should be used to have $\tilde{0}$-presheaf as the lowest value of the energetic spectrum.

The above (partial) realization of the idea that quantum non-commuting observables can commute in some topos, shows that to have commutativity of $X$, $P$ which is connected with the exterior $XP$, one should consider the symmetry of categories involved in the construction. Thus, in a strict sense this is difficult to achieve. Weakening via categorical symmetries is needed. However, (S1) and (S2) are rather strong conditions.

Another construction of the categorical ,,gauge'' symmetries is possible which can give weak categorical agreement between the on-stages interpretations of self-adjoint operators and external products of these interpreted in the topos. The commutativity holds true in the topos up to some categorical gauge. This requires, however the internalisations in some boolean topoi. Taking the isomorphisms of the categories of these topoi as gauge transformations of the interpreted theory, the weak commutativity indeed holds in such a modified spectral topos.   

The basic ingredients of the construction are Takeuti's topoi and boolean valued analysis \cite{takeuti1978}. 

Given commuting a set of self-adjoint operators $\{A_{\alpha}\}$ on some separable Hilbert space ${\cal{H}}$ one can always find the complete boolean algebra $B_A$ of projections such that these projections determine the spectral families of the operators \cite{takeuti1978}. The same holds true for the sets of self-adjoint operators $V_{sa}$ corresponding to every $V\in {\cal{V}}$. The algebra assigned is $B_V$. 

Every automorphism of $V_{sa}$ determines some automorphism of $V$. 

Let us take the boolean topoi $Sh(B_V)$.  
 
Alternatively, one can take the spectra of the unital boolean algebras $\underline{\Sigma}_V$, $V\in {\cal{V}}$. The measure algebras generated by these, as the quotient of the ring of all measurable complex functions on $\underline{\Sigma}_V$ by the ideal of measure zero functions, is the boolean algebra - measure algebra $\tilde{B}_V$. These two algebras are isomorphic \cite{takeuti1978}. Again, let us take the topoi of sheaves of sets on these measure algebras, i.e. $Sh(\tilde{B}_V)$.  

The standard result of the Takeuti's theory is that these boolean topoi are boolean models for Zermello Frankel set theory with the axiom of choice (ZFC), such that real numbers in the model based on measure algebra correspond uniquely to measurable real functions on $\underline{\Sigma}_V$. The real numbers in the previous topos, based on the algebra of projections, correspond uniquely to the self-adjoint operators whose spectral families are built of the projections. 

Thus, the boolean topos $Sh(B_V)\simeq Sh(\tilde{B}_V)$ is assigned to every $V\in {\cal{V}}$. The net of internalisations emerge: on every $V\in {\cal{V}}$, there corresponds the topos $Sh(B_V)$ and one takes the internal reals. As the result, one has internal real numbers instead of self-adjoint operators.  

Let us take internal reals $l_V(\delta(A^2))=l_{1,V}$, $l_V({\delta_V}^2(A))=l_{2,V}$ in $Sh(B_V)$, corresponding to $\delta_V(A^2)$ and $\delta_V(A)\delta_V(A)$ respectively. Create the internal modules of differences of reals $|l_{1,V}-l_{2,V}|=l_V$. 
Let the following relation holds
\begin{equation}\label{sup1}
0\leq l_{1,V}\leq l_{2,V}, \,\, V\in {\cal{V}}  
\end{equation}
and take $d_V=1/2l_V+l_{1,V}$.

Consider the automorphisms $l_{d,V}: R_V\to R_V$ of internal reals, given by $x\to x-d_V$ on each $V\in {\cal{V}}$ where $x$ is an internal real from $R_V$.

Compose each automorphism as above with the automorphism $-1*:R_V\to R_V$ given by the multiplication by $-1$ on every $V\in {\cal{V}}$: $-1*: x\to -1*x$.

This yields the automorphism: 
\begin{equation*}\label{aut2}
(-1*)\circ {l_{d,V}}(l_{1,V})=l_{2,V} 
\end{equation*}
\begin{equation}
(-1*)\circ {l_{d,V}}(l_{2,V})=l_{1,V}
\end{equation}
of $R_V$. This can be extended to the automorphism of the topos $Sh(B_V)$ \cite{takeuti1978}. 

Consider the minimal category of topoi $K$, among objects of which there are $Sh(B_V)$, $V\in {\cal{V}}$ and morphisms of $K$ are geometric morphisms of the boolean topoi. The family of automorphisms 
\begin{equation}\label{aut1}
F_V:=(-1*)\circ l_{d,V},\,\, V\in {\cal{V}}
\end{equation}
defines the automorphism $F_K$ of the category $K$.
\begin{lemma}
Assuming (\ref{sup1}), the uncertainty principle holds in the spectral topos in a weak form, i.e. up to some automorphism of the category $K$. The internal versions of $\delta_V(A^2)$ agrees with $\delta_V(A)\delta_V{A}$ on every $V\in {\cal{V}}$, up to the gauge $F$.
\end{lemma}
This follows from the relation (\ref{aut2}) and the commutativity on-stages in the topos under the $\delta$-interpretation. $\square$

Considering the automorphisms of $K$ as categorical gauges of a theory, the uncertainty principle holds only up to a gauge in such a theory. This kind of results are possible due to the combination of internal and categorical techniques.  
    
\section{\label{secGR} Gravity in $SET^{{\cal{V}}^{op}}$}

In this section we try to add gravity to the topos $SET^{{\cal{V}}^{op}}$ which means that we want to describe consistently gravity as propagated in the topos. To this end one could try to formulate general relativity (GR) in any topos. However, there exist severe obstructions to formulate full GR in a general topos. First, GR should be formulated constructively, i.e. without referring neither to the axiom of choice (AC) nor the 'excluded middle' law. The language would be the internal language of the topos, hence the logic is intuitionistic one. Second, one has to perform differentiation and/or integration in general topoi to formulate GR. However, a theory of integration, measure and differentiation in general topoi is under development (see e.g. \cite{jackson2006}). This is certainly worth performing as a step in the program of formulating a full theory of gravity in a wider class of topoi extending the class of smooth topoi \cite{moerdijk1991}. 

Following Moerdijk and Reyes \cite{moerdijk1991} a different strategy is possible. To formulate a theory of gravity alone with a theory of, say, harmonic oscillator, or other quantum system, in $SET^{{\cal{V}}^{op}}$, the formal language of the system should be modified such that this allows for the use of variable types according to Feferman \cite{feferman1985}. This means that one can also form new types as sub-types of the existing types, by the use of the set theoretical rule of separation stating in this context that for a formula $\phi$,  $\{x\in S: \phi(x) \}$ is a type provided $S$ be. For example, given reals $R$ as a type we can also have a type $T=\{x\in R: x^{1/3}=0\}$ or given a type term $f: R\rightarrow R$ we can have a type term $G=\{f\in R^R: 2f''-f' + 2=0 \}$ etc.. Again, countable set of variables of any given type is assumed to exist in this language. Moreover, equations between terms of arbitrary types are allowed and the quantification on type-terms, say variables, is possible. For example, one can easily formally say ,,for every manifold'' or ,,for every solution of the equation'' etc..

Thus, we can formulate internally in a topos the gravitational field interacting with energy-momentum sources of a system, which can be classical or quantum. In fact, the description of the gravitational field, given in Einstein's equations, is performed externally. The class of solutions appears then as a type of the theory. In that way one does not need to develop internally a full logical theory of GR. 

We are interested in a model in the specific topos rather than full logical theory of interactions of gravity and a quantum system valid in all topoi. That is why, at this stage, we can place all the nonconstructive objects (those requiring AC and the low of excluded middle) as types of the language.  

Thus, given a quantum system ${\cal{H}}$ in spacetime its gravitational interaction with an external (classical) gravitational field, is described via the energy-momentum tensor $T_{\mu\nu}$ of the system. This semi-classical description is the one which we try to internalize in $SET^{{\cal{V}}^{op}}$.
Thus, the formal language of a system should be augmented by the variable types corresponding to the solutions of the Einstein's equations with the sources $T_{\mu\nu}$. One can also consider the Newtonian limit of this gravitational interactions and then interprets this in $SET^{{\cal{V}}^{op}}$.

To build an interpretation of the theory in a topos, one should assign, in particular, objects of the topos to types. The issue of spacetime object in the topos has to be approached.

However, instead of defining spacetime as a specific object in $SET^{{\cal{V}}^{op}}$ we rather modify the local (external) spacetime structure by this topos. We need to internalize only local spacetime data. The generalized spacetimes modified locally by topoi were poposed in the papers \cite{krol2006a,krol2006,krol2007}. Following this, let us assume that local spacetime smooth manifold's structure is locally modified by the topos $SET^{{\cal{V}}^{op}}$. The meaning of locality is understood as taking place at ,,sufficiently small'' spatial distances which now I leave unspecified.  

{\it We say that a quantum system in spacetime locally modifies the spacetime structure provided the structure is locally modified by the spectral topos $SET^{{\cal{V}}^{op}}$ of the system.}

Thus, at sufficiently small scale, the local external patch ${\mathbb{R}}^4$ of a spacetime manifold, is described as the object $R^4$ in $SET^{{\cal{V}}^{op}}$ where $R$ is the object of real numbers in $SET^{{\cal{V}}^{op}}$. Next the change of local coordinates at the region of modification, is performed internally, according to the base of internal topology of $R^4$ given by some system of isomorphic to $R^4$ objects. In the case of a category of presheaves of sets on a small category, the object of Dedekind's reals is always a constant presheaf \cite{maclane1992}. Thus in the topos $SET^{{\cal{V}}^{op}}$ $R$ is a contravariant functor $R:{\cal{V}}\rightarrow \mathbb{R}$ such that $\mathbb{R}$ is assigned to every stage $V\in {\cal{V}}$ and while taking $V'\subset V$ $R(V)=_{id}R(V')$.
Similarly, ${\mathbb{R}}^4$ is a constant presheaf $R\times R\times R\times R\simeq R^4$.

Local descriptions of spacetime $M^4$ are now given by the isomorphic copies of $R^4$'s in $SET^{{\cal{V}}^{op}}$ and the change of coordinates corresponds to some morphisms between constant presheaves $R^4$. Externally, the change of local coordinates on $M^4$ corresponds to some diffeomorphisms $\phi:{\mathbb{R}}^4\rightarrow {\mathbb{R}}^4$ (between diffeomorphic to ${\mathbb{R}}^4$ open subsets of $M^4$), internally it is represented by a morphism $\tilde{\phi}:R^4\rightarrow R^4$.

We want to describe gravitational interactions on such locally modified spaces. That is why, at scales where local modification of spacetime occurs one should internalize: 
\begin{itemize}
\item[1.] Some object equivalent to $R^4$ as a local patch for a space-object $\tilde{M}$ in the topos.
\item[2.] The intersection of local patches as above.
\item[3.] The change of coordinates.
\item[4.] A metric $g:TM\times TM \rightarrow \mathbb{R}$ in the internal coordinates under internal change of these.
\end{itemize}
The above data give rise to the internal in $SET^{{\cal{V}}^{op}}$ local structure of spacetime which is modified by the topos $SET^{{\cal{V}}^{op}}$. The internal in categories version of spaces and their local covering families were presented by Bartels \cite{bartels2004}. Following this work, to express that $R^4$ is a local chart of some internal space $\tilde{M}$ (assume here, without specifying, that some internal space-object $\tilde{M}$ exists) is equivalent to the existence of the morphism of presheaves $i:R^4\rightarrow \tilde{M}$. The morphism corresponds to the external diffeomorphism $\phi:{\mathbb{R}}^4\rightarrow U\subset M$ such that at every stage $V$ the corresponding morphism in $SET$, $i_V:{\mathbb{R}}^4\rightarrow {\mathbb{R}}^4$, is exactly $\phi$.

To say that two local patches $\tilde{i_1}:R^4\rightarrow \tilde{M}$ and $\tilde{i_2}:R^4\rightarrow \tilde{M}$ of $\tilde{M}$ intersect is to specify the intersaction object $\tilde{i_1}(R^4)\cap \tilde{i_2}(R^4)$ as a pullback in $SET^{{\cal{V}}^{op}}$ of this two morphisms:
\begin{equation}\label{diag1}
\xymatrix{
   \tilde{i_1}(R^4) \cap \tilde{i_2}(R^4)
     \ar[rr]^{\tilde{j_1}}
     \ar[dd]_{\tilde{j_2}}
      && R^4
     \ar[dd]^{\tilde{i_1}} \\ \\
      R^4
     \ar[rr]^{\tilde{i_2}}
      && \tilde{M} }
\end{equation}
In terms of constant presheaves, at every component $V\in {\cal{V}}$, one has the corresponding diffeomorphic change of local coordinates, i.e. $\phi:{\mathbb{R}}^4\rightarrow {\mathbb{R}}^4$ where $(i_{1}\circ {i_{2}}^{-1})_{|V}=({i_{1}})_{|V}\circ ({i_{2}}^{-1})_{|V}=\phi$.

$TM_{|{{\mathbb{R}}^4}}$ is interpreted in local coordinates as a constant sheaf $T_0{\mathbb{R}}^4\simeq {\mathbb{R}}^4$ at every stage $V$, i.e. $R^4$. Any function $g: TM\times TM\rightarrow {\mathbb{R}}^4$ would correspond to a functional symbol of the corresponding signature in the language. Interpreting these in $SET^{{\cal{V}}^{op}}$ and expressing in local coordinates one gets a morphism $\tilde{g}:R^4\times R^4\rightarrow R$ between the constant presheaves. In terms of local coordinates $R^4_1$ this reads $\tilde{g_1}:R_1^4\times R_1^4\rightarrow R$. The change of coordinates as in \ref{diag1}, leads to the internal change of coordinates in a tangent space, given by the following pull-back diagram:
\begin{equation}\label{diag2}
\xymatrix{
   (R^4_1\times R^4_1) \times_{\tilde{TM}\times \tilde{TM}} (R^4_2\times R^4_2)
     \ar[rr]^{\tilde{j_1}}
     \ar[dd]_{\tilde{j_2}}
      && R^4_1\times R^4_1
     \ar[dd]^{d(\tilde{i_1}\times \tilde{i_1})} \\ \\
      R^4_2\times R^4_2
     \ar[rr]^{d(\tilde{i_2}\times \tilde{i_2})}
      && \tilde{TM}\times \tilde{TM}}
\end{equation}
Here $d(\tilde{i_1}\times \tilde{i_1}\circ (\tilde{i_2}\times \tilde{i_2})^{-1})_{|V}=d\phi\times d\phi$ where $\phi:{\mathbb{R}}^4\rightarrow {\mathbb{R}}^4$ was the diffeomorphic change of local coordinates.

The corresponding change of metric, when expressed in local coordinates, can also be written down component-wise in the topos $SET^{{\cal{V}}^{op}}$. 

Having expressed local change of coordinates in $SET^{{\cal{V}}^{op}}$ and the corresponding change of metric one can similarly express local changes of various objects written down in local coordinates on spacetime. All types needed in the language as signatures of functional symbols, can be generated outside as variable types of the theory. 

Now we try to interpret in the topos also global information about spacetime. 
The type of some class of solutions $Sol_{EE}$ of the Einstein's equations with a suitable stress-energy terms can be generated. Thus, the corresponding variables of this type, when interpreted in a topos as some presheaves, encode in the topos also global information of the solutions. Let us assume that spacetime $M^4$ is determined by a metric $g:TM\times TM \rightarrow \mathbb{R}$ which is in the allowed family of solutions of the Einstein's equations $G_{\mu\nu}+\Lambda g_{\mu\nu}=8\pi\chi T_{\mu\nu}$ and $T_{\mu\nu}$ is some stress-energy tensor. This family can be considered as a type $Sol_{EE}$ of our theory. Thus, $T_{\mu\nu}$ is fixed in the type. To express in the variable types theory that one has a specific solution can be formulated as $\exists_xF(x)$ where $x$ is a variable of type $Sol_{EE}$ and $F(x)$ is a formula (a conjunction of equations) which gives specific bounds on the solution $x$. 
 
Thus, a spacetime $M$ can be interpreted in $SET^{{\cal{V}}^{op}}$ as follows:
\begin{itemize}
\item[1.] $\tilde{M}$ is an object which can be constant $M^4$ on every stage $V$. $\tilde{M}$ is an individual constant of a type $Sol_{EE}$ and is interpreted by an arrow $\tilde{M}:1\rightarrow_{M^4} Sol_{EE}$. One could alternatively assume that $M^4$ were the type symbol and countably many variables of this type would exist in the language. In this case $\tilde{M}$ would be a variable of type $M^4$. This would be important for more global and dynamical {\it internal} description of gravity in the topos. 
\item[2.] Local external charts $i_1,i_2:{\mathbb{R}}^4\rightarrow U_1,2\subset M^4$ of a smooth manifold $M^4$ where $i_1,i_2$ are diffeomorphisms and $i_1\circ i_2^{-1}:{\mathbb{R}}^4\rightarrow {\mathbb{R}}^4$ is also a diffeomorphism, are interpreted in $SET^{{\cal{V}}^{op}}$ as $\tilde{i_1},\tilde{i_2}$ which are maps between presheaves as in the diagram above, and are generated by constant on every stage $V\in {\cal{V}}$ smooth maps $i_1, i_2$.
\end{itemize}
As the result of this modification of the language of a quantum system by variable types and interpretation of the local structure of spacetime in the spectral topos of the system, and corresponding interpretation of metric, Christoffel symbols and the like, one can express the way how gravity interacts with a quantum system in spacetime.

At scales where gravity and the quantum system are both described internally in $SET^{{\cal{V}}^{op}}$, the topos modifies both, the description of a quantum system and the local structure of spacetime.

{\it A quantum system ${\cal{H}}$ in spacetime interacts with an external gravitational field via the spectral topos determined by the system. The local spacetime structure is modified by the topos.}
 
The following points give more detailed content of the above rule:
\begin{itemize}
\item[1.] The local structure of spacetime is internally described with respect to the object $R$ of real numbers in $SET^{{\cal{V}}^{op}}$ (i.e. the constant presheaf $R$ in $SET^{{\cal{V}}^{op}}$). 
\item[2.] Gravity is represented locally by the internal metric $\tilde{g}$ and Christoffel's symbols (and curvature tensor) on the internal 4-manifold $\tilde{M}$. These are also constant presheaves.  
\item[3.] Quantum observables of the quantum system are represented by on-stages (outer) daseinisation presheaves. The semi-classical interpretation of observables by morphisms $\underline{\Sigma}\rightarrow R_{\cal{H}}$, is valid up to the multiplicative structure of observables. This is build with respect to the quantity-value object $R_{\cal{H}}$ which is not any constant presheaf in $SET^{{\cal{V}}^{op}}$.  
\item[4.] The topos modifies both, the description of a quantum system and local structure of spacetime. 
\item[5.] The Hamiltonian of the system is $\delta_0$-interpreted in the topos. The internal energetic spectrum contributes to the gravitational interactions via its energy-momentum tensor which acts as external gravitational sources. 
\item[6.] The model of interactions modifies the quantum vacuum energy of the system for the external observer, since $\mathbb{R}\ni 0\mapsto I(0)=\tilde{0}\in R$ where $I:{\cal{L}}\rightarrow SET^{{\cal{V}}^{op}}$ is the interpretation of the language ${\cal{L}}$. This means that $\tilde{0}$ is the global element of the presheaf $R$. 
\end{itemize}
Results of physical experiments are canonically formulated in terms of real numbers $\mathbb{R}$, i.e. constant presheaves of reals in $SET^{{\cal{V}}^{op}}$. Thus, the direct contribution, derived from the internal spectrum, which modifies the external spectrum of a harmonic oscillator, is 0.

This follows from Lemma \ref{lemma3} and the fact that {\it $0\in \mathbb{R}$ is interpreted in $SET^{{\cal{V}}^{op}}$ as the constant presheaf $\tilde{0}\in {R}$.}

Finally, we have 
\begin{theorem}\label{th1}
A quantum system modifies locally smooth manifold spacetime structure such that the uncertainty principle does not hold and energy 0-modes of the system are allowed. These give vanishing contributions to the cosmological constant. 
\end{theorem}
Let us mention that other on-stages internal values of the energy spectrum are not constant presheaves. These are not preasheaves in general. That is why these values can not appear, as they stand, as outcomes of external experiments. 

This theorem holds true, however, under the $\delta_0$-interpretation for $H$. We set as $0$ those $H_V$ which are obtained from the operators $\delta_V(P)$, $\delta_V(X)$ not having $0$ in their spectra. The motivation doing this was to consider $P_V$ and $X_V$ for different $V\in {\cal{V}}$ as approximations of ,,true'' $P$, $X$. This agrees with daseinisations of self-adjoint operators which are best possible approximations on-stages of these operators. True $P$, $X$ have $0$ in their spectra. The approximations preserve this value on-stages. Thus, the result follows at this level of approximation. 

What can happen when $H_V$'s as above are not set to zero? Let us switch on again these non-zero lowest values for $H_V$ which were nullified by the $\delta_0$-interpretation of $H$. The appearance of non-zero eigenvalue for $H_V$ at a stage $V$, as the lowest one, physically would mean that some potential appeared. In general, we are not allowed to consider the on-stages results as physical, however when considering these as approximations of what can happen in full classical or intuitionistic logic, this might be reasonable. These on-stages potential barriers, to have a classical meaning, should be approximated by some constant presheaf in the topos. In the internal intuitionistic case, some, not necessarily constant, presheaf approximation would be required. Thus, assume that there exists a classical approximation for the non-zero on-stages potentials, which is some non-zero constant presheaf. Non-zero minimal values of energy of an internal quantum system, appear. What physical meaning can have these ,,corrections''? We claim that this kind of effects, when considering composite quantum systems and fields in spacetime, can correspond to the appearance of non-zero, small value of the cosmological constant in spacetime. The analysis of this issue, however, is not our concern here and will be presented in the separate paper. 

Topos $SET^{{\cal{V}}^{op}}$ can be construed whenever the Hilbert space ${\cal{H}}$ of a system exists, hence, one is not confined entirely by the case of a quantum oscillator, rather any quantum system defined via its Hilbert space is allowed. Thus, the result in Theorem \ref{th1} is quite general. However, still in the case of many quantum systems in spacetime, each modifying locally the spacetime structure, the cancellation of the contributions from all of them, requires some additional structure. In fact, a category of systems and authomorphisms of it are relevant here. However, this gives rise directly to higher groups. 

\section{Higher Symmetry Group and Gravity}\label{2GR}
In this section I want to discuss sketchily the emergence of higher symmetry groups in the model of interaction of gravity and quantum systems in spacetime, and some consequences of this. The very important issue like that will be presented in details in the forthcoming paper \cite{krol2008a}.

Many choices have been made up to now, which may break covariance of the theory of gravitational interactions with systems in spacetime. In order to free the theory of the choices made one should recognize carefully the symmetry pattern emerging here since the spacetime structure was also modified. In fact, the meaning of diffeomorphisms invariance of the theory has to be enhanced in a suitable categorical sense.  

Assuming that gravity interacts with a quantum system in spacetime via its topos and given many systems, each modifying spacetime locally, one has to consider a category of systems and its action on spacetime. One such category of systems was given by D\"oring and Isham \cite{isham2007d}. Assigning the spectral topoi to quantum systems, the category $TOP_{s}$ of spectral topoi and geometric morphisms between these emerges \cite{isham2007d}. The topoi modify the local structure of spacetime. Given classical systems in spacetime the topos assigned to them is constantly $SET$ and the spacetime structure is not modified. Thus, in the case of quantum systems in spacetime one has to relate the change of local spacetime coordinates with the change of topoi in the category $TOP_s$ since local coordinate charts can be internal in different topoi. Thus it emerges a kind of modified ,,smooth'' structure of such spacetime modified by topoi. 

We saw in Sec. \ref{secIE} that the connection between true $XP$ and $PX$ interpreted in $SET^{{\cal{V}}^{op}}$ and these given by the $\delta$-interpretation, can be recovered up to some automorphism $F$ of the spectral topos, under the suppositions (S1) and (S2).

Whenever system modifies spacetime, local gravitational field is defined internally in $SET^{{\cal{V}}^{op}}$ and interacts with internal energy modes. What survives for the external observer should be expressible in constant presheaves corresponding to the external real numbers $\mathbb{R}$. This scenario requires, however, a kind of gauge freedom which has to be categorical. An additional internal\footnote{internal as in gauge theory: not purely spacetime symmetry.} symmetry of such defined model for gravitational interactions would be connected with a kind of authomorphisms group of the category of topoi $TOP_s$ assigned to the systems in spacetime. However, taking into account the automorphisms of the spectral topoi, the automorphisms of the category of these give rise to the structure of some 2-group rather than a group \cite{baez2005}.  

Next step would be the inclusion a continuum limit of the discrete category $TOP_s$ and its automorphisms. This should be done in order to include also quantum fields. The continuum limit of $TOP_s$ is also crucial for the categorical connection with 2-Yang-Mills theories which naturally contains some string theory data \cite{kapustin2000}. Alone with the continuum limit the structure of some Lie groups appear. When taking suitable automorphisms of these Lie groups, considered as a category with one object, some 2-Lie-group structures emerge. These last are basic ingredients of a 2-Yang-Mills theory. The detailed analysis of this interesting point as well as higher categories constructions involved here is the topic of the forthcoming paper.

It is known that group of automorphisms of any category respecting also isomorphisms of its objects, is a 2-group \cite{baez2004}. Objects of this 2-group are autoequivalences of the category of topoi and natural transformations (isomorphisms) of these isomorphisms are 2-cells \cite{baez2004}. Autoequivalence is understood as a reversible functor $F:TOP_s\to TOP_s$.

Our aim would be the description of the invariance of gravity with respect to this 2-group.

Given a quantum system ${\cal{H}}_1$ and the corresponding topos $\tau_1 =SET^{{\cal{V}}^{op}}$ in $TOP_s$, one assigns the topos to an open subset $U_1\subset M^4$ of the spacetime $(M^4,g)$. Gravitational interactions with the system are described then by the interpretation of $U_1\simeq_{diff} {\mathbb{R}}^4$, $g|_{U_1}$, and related quantities like Christoffel symbols, in $\tau_1$.

Given another system ${\cal{H}}_2$ in spacetime, let its topos, $\tau_2$, be assigned to some open subset $U_2\simeq_{diff} {\mathbb{R}}^4$ of $M^4$. If the topoi of two systems ${\cal{H}}_1$ and ${\cal{H}}_2$, i.e. $\tau_1$ and $\tau_2$, are related by some geometric morphisms, their images under autoequivalence, stay related by the geometric morphisms which are images by this autoequivalence of the initial geometric morphisms. We require that no new local gravitational effects appear in spacetime due to the action of the autoequivalences as above. Moreover, we would like to be sure that no new local gravitational effects in spacetime occur due to the action of the natural isomorphisms on autoequivalences.

{\it The theory would be functorially invariant with respect to the 2-group $AUT(TOP_s)$ provided no new local gravitational effects appear in spacetime $M^4$ while changing local charts in $M^4$ which are generated only by changes of topoi in $TOP_s$. The changes of topoi correspond to autoequivalences and 2-natural isomorhisms in $AUT(TOP_s)$.}

New local gravitational effects are understood comparing these with obtained via standard GR on $M^4$.

To be sure no new gravitational effects appear, the change of local charts in $M^4$ which were chosen to be Minkowski flat and the transformations of these driven by the changes of topoi and morphisms by autoequivalences and natural isomorphisms, should correspond to Lorentz transformations of the local coordinates. 

In other words, local chart whose tangent space is Minkowskian remains Minkowskian after the ,,change of coordinates'' corresponding to a change of topoi in $TOP_s$ which are generated by $AUT(TOP_s)$. 

The 2-group structure of $AUT(TOP_s)$ can be described by the diagrammatic equations:
\begin{equation}\label{diag3}
\xymatrix{
   TOP_s \bullet \ar@/^2pc/[rr]^{F_1}_{}="0"
           \ar[rr]^<<<<<<{F_2}_{}="1"
           \ar@{=>}"0";"1"^{2^{\tau}_1}
           \ar@/_2pc/[rr]_{F_3}_{}="2"
           \ar@{=>}"1";"2"^{2^{\tau}_2}
&&\bullet TOP_s 
}
=
\xymatrix{
   TOP_s \bullet \ar@/^1pc/[rr]^{F_1}_{}="0"
           \ar@/_1pc/[rr]_{F_3}_{}="1"
           \ar@{=>}"0";"1"^{{}_{2^{\tau}_2\circ 2^{\tau}_1}}
&&\bullet TOP_s
}
\end{equation}

\hfill\break
\vskip 1em \noindent
\begin{equation}\label{diag4}
\xymatrix{
   \bullet \ar@/^1pc/[rr]^{F_1}_{}="0"
           \ar@/_1pc/[rr]_{F_1^\prime}_{}="1"
           \ar@{=>}"0";"1"^{2^{\tau}_1}
&& \bullet \ar@/^1pc/[rr]^{F_2}_{}="2"
           \ar@/_1pc/[rr]_{F_2^\prime}_{}="3"
           \ar@{=>}"2";"3"^{2^{\tau}_2}
&& \bullet
}
 =
\xymatrix{
   \bullet \ar@/^1pc/[rr]^{F_1 \circ F_2}_{}="0"
           \ar@/_1pc/[rr]_{F_1^\prime\circ F_2^\prime}_{}="1"
           \ar@{=>}"0";"1"^{{}_{2^{\tau}_1 \circ 2^{\tau}_2}}
&& \bullet 
}
\end{equation}
\hfill\break
\vskip 1em \noindent
These should be reflected by the corresponding nets of Lorentz transformations between Minkowski flat local patches, assuming that all corresponding changes of local patches are governed by changes of topoi as in the 2-group $AUT(TOP_s)$. 

\begin{equation}\label{diag5}
\xymatrix{
 U_{\tau_1} \ar@/^2pc/[rr]^{\Lambda_1}_{}="0"
           \ar[rr]^<<<<<<{\Lambda_2}_{}="1"
           \ar@{=>}"0";"1"^{\Lambda_4}
           \ar@/_2pc/[rr]_{\Lambda_3}_{}="2"
           \ar@{=>}"1";"2"^{\Lambda_5}
&& U_{\tau_2}
}
=
\xymatrix{
   U_{\tau_1} \ar@/^1pc/[rr]^{\Lambda_1}_{}="0"
           \ar@/_1pc/[rr]_{\Lambda_3}_{}="1"
           \ar@{=>}"0";"1"^{{}_{\Lambda_5\circ \Lambda_4}}
&& U_{\tau_2}
}
\end{equation}

\hfill\break
\vskip 1em \noindent
\begin{equation}\label{diag6}
\xymatrix{
   U_{\tau_1} \ar@/^1pc/[rr]^{\Lambda_1}_{}="0"
           \ar@/_1pc/[rr]_{\Lambda_1^\prime}_{}="1"
           \ar@{=>}"0";"1"^{\Lambda_4}
&& U_{\tau_2} \ar@/^1pc/[rr]^{\Lambda_2}_{}="2"
           \ar@/_1pc/[rr]_{\Lambda_2^\prime}_{}="3"
           \ar@{=>}"2";"3"^{\Lambda_5}
&& U_{\tau_3}
}
 =
\xymatrix{
   U_{\tau_1} \ar@/^1pc/[rr]^{\Lambda_1 \circ \Lambda_2}_{}="0"
           \ar@/_1pc/[rr]_{\Lambda_1^\prime\circ \Lambda_2^\prime}_{}="1"
           \ar@{=>}"0";"1"^{{}_{\Lambda_4 \circ \Lambda_5}}
&& U_{\tau_3} 
}
\end{equation}
%\hfill\break
%\vskip 1em \noindent 
Where all $\Lambda_i,i=1,...,5$ are some Lorentz transformations of $T_0U\simeq U$ and all $T_0U_{\tau_i}\simeq U_{\tau_i}$ are stated up to some Lorentz transformations.

In the smooth limit of this 2-category data with smooth assignment of Lorentz group elements, the requirement that we are confined by some Lorentz transformations, when changing internal patches by $AUT(TOP_s)$, is possible only for some abelian subgroup of the Lorentz group. This follows from the fact that when one tries to decorate consistently the diagrams \ref{diag3} and \ref{diag4} regarded as smooth paths and surfaces by some smooth group elements, the group has to be abelian \cite{baez2004}. If not, one has to take some 2-group as decorating group \cite{baez2004}. 

Similarly, taking diffeomorphisms of ${M}^4$, as decorating group for the smooth version of the diagrams \ref{diag3} and \ref{diag4}, we deal with some abelian subgroup of Diff, or we allow for the extension of symmetries to some 2-group \cite{baez2004}.

In the case of this extension, the ,,change of local coordinates'' modifies the usual changes given by diffeomorphisms. The theory is not necessary invariant with respect to the extended changes of local charts corresponding to all geometric morphisms between spectral topoi from $TOP_s$. 

Some gravitational effects in spacetime could appear due to such change of topoi and local coordinates. From the other side, the patches are internal objects in topoi, thus they cannot be all valid choices for coordinate patches generating or excluding the effects of gravity, as given by the equivalence principle of GR. Even though the local effects are excluded, we claim that this has global manifestation. The global effects are understood, however, in the sense of higher Yang-Mills theory which is based on some 2-group. These are higher curvature and higher holonomies. 
Gravity, when 2-symmetric in modified spacetime could be described by some higher Yang-Mills theory.

Such a higher YM theory can relate ordinary spacetime bosons with fermions. The emerging supersymmetry is worth considering as derived structure which appear in non-supersymmetric fundamental theory.

The construction and details of the argumentation will be presented in the forthcoming paper. Summarizing, we have the theorem: 
\begin{theorem}
1. Any quantum complex system of quantum oscillators in spacetime which are represented by the spectral topoi from the category $TOP_s$, with the $\delta_0$-interpretation for $H$ and $\delta$-interpretation for its quantum observables, has vanishing zero-points contributions to the cosmological constant. The fundamental symmetry group of the emerging categorical theory of gravitational interactions with such systems, is the smooth version of the 2-group $AUT(TOP_s)$. 

2. Some global gravitational effects in spacetime appear due to the generalized change of internal coordinates driven by $AUT(TOP_s)$. The effects cannot be localized in spacetime.
\end{theorem}

To have local YM-type theory one has to take a smooth limit of 2-group $AUT(TOP_s)$. This can be done via internalisation in some smooth categories \cite{baez2005}. This parallels the continuum limit where quantum fields can be discussed. The emerging symmetry is a local gauge 2-symmetry of the theory in the sense of higher Yang-Mills theory of Baez at all. \cite{baez2004}. In the abelian limit of such a theory, one can recover, as the higher connection coefficients, the fields from the Neavou-Schwarz sector of the IIB sustring theory compactified to 4-dim \cite{kapustin2000}.

The effects of the higher symmetries cannot be localized in spacetime, since internal 4-patches in the topoi are not true coordinate patches for the observer unless gravity is strong at small scales. Only gravity at this scale recognizes internal in topoi patches, since it is propagated in the topos and is higher symmetric. If gravity forces would be dominant over other forces at small distances the higher symmetry would be also evident for an observer. Gravity dominates at large scales but the modification of spacetime takes place at small distances. 

The correct recognition of the higher symmetry of gravity and spacetime seems to be crucial for approaching the problem of quantum gravity. The discussion of this relation will be also included in the forthcoming paper. 

\section{Concluding Remarks}\label{remark}
I presented the scenario which is a way toward the cancellation of the contributions to cc of the energy modes of any quantum system in spacetime. The important thing in the construction was the representing spacetime and gravity with respect to the object of real numbers in $SET^{{\cal{V}}^{op}}$. This object is a constant presheaf. Thus, all gravity and spacetime related quantities can be represented tautologically by constant presheaves. This simplifies the problem of the interpretation. In contrary, a quantum system becomes semi-classical with respect to observables corresponding to the morphisms between presheaves, but with respect to the constructed by D{\"o}ring and Isham quantity-value object $R_{\cal{H}}$. This is not a constant presheaf and is not an object of real numbers in $SET^{{\cal{V}}^{op}}$. There is a big gap between the two objects. However, I did not make any essential use of the $R_{{\cal{H}}}$ object in the paper, since the interpretation of the language was modified such that products of quantum operators can be also interpreted as on-stages products of daseinisations in the topos. The connection with the semi-classical interpretations of quantum observables by morphisms in the topos for products of the observables is lost.     
This indicates on some higher gauge symmetry emerging here, which partially recovers the connection.

The local modification of spacetime structure by local charts internal in topoi, suggests a somewhat different approach to the problem of quantity-value object which would make a quantum system semi-classical. When taking topoi of sheaves on maximal boolean subalgebras of the lattice of projections $\mathbb{L}_{\cal{H}}$, the object of reals can be replaced by a locally boolean object of reals. This object is in a half way between constant sheaf $R$ and quantity-value object $R_{\cal{H}}$. Spacetime and gravity can be described with respect to this object and locally boolean reals can be considered as the quantity-value object for the quantum observables. This description via Takeuti's topoi will be presented in a separate paper where also higher structure, like gerbes and stacks, will be important (see, however, \cite{krol2007}). The approach to physics via model theory and Takeuti's models of set theory (Takeuti's topoi and boolean-valued analysis) goes back to works by Takeuti, Scott and Benioff (see e.g. \cite{takeuti1983,benioff1976}). This approach is still present in physical writings and is under development.

More internal, than merely by constant sheaf, developing a theory of gravity in topoi, is worth performing, especially when quantum aspects of gravity become relevant. This is connected with general idea of considering physics as intuitionistic rather than classical at the fundamental level \cite{krol2005,krol2006a}. 

The draft of the description of higher symmetry of gravity given in Sec. \ref{2GR} will be presented in details in the forthcoming paper \cite{krol2008a}. The emerging higher symmetry of gravity and spacetime is crucial for the correct recognition of the cancellation mechanism and for the approaching gravity also at the quantum level (cf. \cite{krol2007}). The very attractive possibility is the relation of any {\it smooth} 2-group (or strict Lie 2-group) with gerbes via the connection coefficients of some locally trivial 2-bundle with this structure 2-group. This has direct relation with the description of branes in string theory.  

Even though the contributions of the density of energy of quantum vacua can disappear, the small non-zero value of the effective cosmological constant should be somehow explained in the presented framework of categories and topoi. We do not approach this issue directly in this paper, however the cancellation of the contributions was achieved supposing the $\delta_0$-interpretation for $H$ of a quantum oscillator. Without this, there appear non-zero contributions. As was observed in Sec. \ref{secGR} this can be responsible for small non-zero value of the effective cc. To have some quantitative results one should consider the on-stages $<H_V>$ as approximations to the external quantum mechanical value $<H>$. On these approximations some probability measure can be defined. The measure should be higher symmetric invariant and has its gaussian maximum for those $H_V$ which correspond to $P_V$ and $X_V$ possessing zeros in their spectra. These are, however, nullified by the commutativity. Remaining non-zero contributions can be done sufficiently small with respect to the measure.     

From the other side the issue of non-zero small value of the effective cc seems to be connected with exotic smooth 4-structures in spacetime. The proposition that exotic 4-smoothness on spacetime manifold can be responsible for the cc term (dark energy) was considered by Brans \cite{brans1994}, S{\l}adkowski \cite{sladkowski1999,sladkowski2001} and Asselmeyer and Brans \cite{BransAsselmeyer2002}. The realization of this program alone with the calculation of the value of the cc, based on the interplay of exotic smoothness in 4 dimensions and 3-dimensional geometry and topology, was performed by Asselmeyer and Ros\'e \cite{asselmeyer2005,asselmeyer2006}. There exists also the proposal to relate exotic 4-smoothness on open 4-manifolds with the local modification of the smooth 4-manifolds by topoi \cite{krol2005}. From that point of view it is quite natural to look for the explanation of the non-zero value of cc by emerging exotic smooth structures. However, at present the relation of topoi and smooth exotic open manifolds is not rigorously formulated and we postpone this interesting issue for future work.

Even more natural possibility to generate non-zero cc would be via a kind of breaking the higher symmetry of gravity. The discussion of the scale at which it can happen and the experimental bounds related with this mechanism are certainly worth performing. Breaking of higher symmetry of gravity would be also related with the categorical modification of spacetime, thus, possibly, with the exotic smoothness again. 

The approach to the cosmological constant problem via topoi and (higher) categories, presented in this paper, is new one and this is the first paper relating this issue. More work is needed as e.g. on the possible experimental bounds on the proposed modification of gravity and spacetime or the careful recognition of the symmetry breaking pattern and the relation with quantum gravity. Certainly, the continuum limit of the higher symmetry is crucial for the QFT limit of the approach as well as for the relation with some sector of string theory. 

The proposed approach does not make any essential use of supersymmetry. Rather a fundamental symmetry is a kind of higher symmetry of gravity and generalized spacetime but not of the theory of quantum systems. This last theory is modified toward pre-semi-classical or semi-classical theories.
\subsection*{Appendix A}\label{app1}
\subsubsection*{The formal language ${\cal{L}_{H}}$, its representation  in $SET^{{\cal{V}}^{op}}$ and the extension by variable types}
\begin{itemize}
\item Set of symbols.
\subitem{1.} Basic type symbols i.e. $1,\Omega,\Sigma,{\cal{R}},{\mathbb{R}}$. One should use the usual recursive rules for the creating new type symbols from the existing already (power-type and Cartesian product of types). 

When interpreting in the spectral topos, $1$ becomes a terminal object of the topos, $\Omega$ is the subobject classifier $\underline{\Omega}$ of the topos, $\Sigma$ becomes a states object i.e. the spectral presheaf $\underline{\Sigma}$ \cite{isham2007a}, ${\cal{R}}$ becomes the presheaf $R_{{\cal{H}}}$, i.e. ,,quantity-value'' object \cite{isham2007c} and $\mathbb{R}$ becomes the constant presheaf of real numbers in the topos, denoted by $R$. Any type $T$ is represented by some object, i.e. presheaf $I(T):=\underline{T}$, in the topos.
\subitem{2.} For each type symbol there exist a countable set of variables of this type.
\subitem{3.} The special symbol $\ast$.
\subitem{4.} A set $F_{{\cal{L}_H}}(T_1,T_2)$ of function symbols $A:T_1\rightarrow T_2$, is assigned to each pair of types $(T_1,T_2)$. Some of these sets can be empty. 

Every function symbol $A:T_1\rightarrow T_2$ of signature $(T_1,T_2)$ is represented by a morphism between presheaves $\hat{A}:\underline{T_1}\rightarrow \underline{T_2}$.
\item Set of terms.
\subitem{1.} The variables of type $T$ (the existence of countably many of these is assumed in 2. above) are terms of type $T$.

Every variable $t$ of type $T$ is represented also by $\underline{T}$. Every constant $c$ of type $T$ is a global element i.e. a morphism $1\rightarrow \underline{T}$.
\subitem{2.} $\ast$ is a term of type $1$.
\subitem{3.} A formula is any term of type $\Omega$. When a formula has no free variables it is a sentence.
\subitem{4.} Given a function symbol $F:T_1\rightarrow T_2$ and a term $t$ of type $T_1$, the term $F(t)$ has type $T_2$.
\subitem{5.} For terms $t_1,t_2,...,t_n$ of types $T_1,T_2,...,T_n$, the term $<t_1,t_2,...,t_n>$ has type $T_1\times T_2\times ...\times T_n$.

Every term $t(\overrightarrow{x})$ of type $T$ with free variables among $\overrightarrow{x}=(x_1,...,x_n)$ which are of types $\overrightarrow{t}=(T_1,...,T_n)$ is represented by a morphism $\underline{\overrightarrow{t}}\rightarrow \underline{T}$ between the corresponding presheaves.
\subitem{6.} Given $t$ of type $T_1\times T_2\times ...\times T_n$, the term $(t)_i$ is a term of type $T_i$, for $1\leq i \leq n$. 
\subitem{7.} For a term $\phi$ of type $\Omega$ (a formula) and a variable $v$ of type $T$, $\{v:\omega \}$ is a term of power type of the type $T$, i.e. $PT$.

The power type $PT$ is represented by a power sheaf $P\underline{T}$ of a sheaf $\underline{T}$.
\subitem{8.} $t_1=t_2$ is a term of type $\Omega$, provided both $t_1$ and $t_2$ are terms of the same type.
\subitem{9.} Let $t_1$, $t_2$ be terms of type $T$, $PT$ respectively. $t_1\in t_2$ is a term of type $\Omega$.
\subitem{10.} A sentence $\omega$ i.e. a term $\omega$ of type $\Omega$ with no free variables, is represented by a morphism (a global element of $\underline{\Omega}$ = a truth value) $1\rightarrow \underline{\Omega}$.
\item Axioms and rules of inference.
\subitem{11.} Usual axioms and rules of inference of predicate intuitionistic calculus are assumed \cite{lambek1994}.
\subitem{12.} The specific axioms regarding ring of real numbers or natural numbers are included. Some axioms which can be motivated by physics are allowed, however it requires much care to consider some property as logically valid \cite{isham2007a}.
\item A class of additional symbols, like parenthesis, is usually allowed.
\end{itemize}

The above higher order, type intuitionistic language is rather rigid and weak, especially when one wants to express in it various sentences regarding analysis on manifold or GR etc.. That is why, in Sec. \ref{secGR} we made use of the theory with variable types of Feferman \cite{feferman1985}. The types of such a theory are generated by arbitrary conditions/equations. The adventage of this is that types are built outside, at the level of language symbols, i.e. before interpreting as objects of a topos. Still, the logic assumed is the intuitionistic one, hence the interpretation in topoi is allowed. 

The formal language of variable type theory has to be augmented by a class of type terms denoting types. One can, in particular, quantify over these terms, hence over types. Moreover, equations between terms of different types are allowed.
\begin{itemize}
\item[1.] Variables $X$, $Y$ are type terms.
\item[2.] If $T_1$ and $T_2$ are type terms, $T_1\times T_2$ and $T_1\rightarrow T_2$ are either.
\item[3.] If $\phi$ is a formula, then $\{x\in T_1: \phi \}$ is a type term.
\item[4.] An equality $t_1=t_2$ is allowed even the terms are of different types.  
\end{itemize}
One can also specify type constants, like $1$, $N$, $R$, $\Omega$ and individual constants of a given type, like $0$, $1$, $+$, $\cdot$, $\leq$ of types $R$, $R$, $R\times R\rightarrow R$, $R\times R\rightarrow R$, $R\times R\rightarrow \Omega$, etc.. The specified type constant refer to the singleton, natural numbers, real numbers and the truth values. 

Equipped with such a way to built types, one can easily express in the formal language many needed components, like topology of, say, $R$: it is a subtype ${\cal{O}}(R)$ of the power type $PR$ or it is the type term denoting the order topology of $R$ which is defined via the order relation $<$ \cite{moerdijk1991}.
Every self-adjoint operator on a Hilbert space ${\cal{H}}$ is interpreted by on-stages daseinisations of the operators. The products of operators are interpreted again as the on-stages products of daseinisations.  
 
\subsubsection*{Acknowledgements}{I thank Pawe{\l} Gusin for fruitful and long discussions of the matters from the paper.}
\subsection*{\ \ References}

\newcounter{bban}

\begin{list}
{[\arabic{bban}]}{\usecounter{bban}\setlength{\rightmargin}
{\leftmargin}}
\bibitem{BransAsselmeyer2002} T. Asselmeyer-Maluga, C.H. Brans, ''Cosmological anomalies and exotic smoothness structures,'' {\it Gen. Rel. Grav.} {\bf 34}(10) (2002).
\bibitem{asselmeyer2006} T. Asselmeyer, H. Ros{\'e}, Calculation of the cosmological constant by unifying matter and dark energy, {\tt gr-qc/0609004} (2006).
\bibitem{asselmeyer2005} T. Asselmeyer, H. Ros{\'e}, Differential structures-geometrization of quantum mechanics, {\tt gr-qc/0511089} (2005).
\bibitem{baez2004} J.C. Baez, U. Schreiber, Higher gauge theory: 2-connections on 2-bundles, {\tt hep-th/0412325} (2004).
\bibitem{baez2005} J.C. Baez, U. Schreiber, Higher gauge theory, in {\it Categories in Algebra,
Geometry and Mathematical Physics}, eds. A. Davydov et al, {\it Contemp. Math.} {\bf 431}, AMS, Providence, Rhode Island, 2007, pp. 7–30, {\tt math.DG/0511710} (2005).
\bibitem{baez1998} J.C. Baez, J. Dolan, Categorification, in {\it Higher Category Theory}, eds. Ezra Getzler and Mikhail Kapranov, {\it Contemp. Math.} {\bf 230}, AMS, Providence, Rhode Island, (1998), pp.1-36, {\tt math/9802029}.
\bibitem{bartels2004} T. Bartels, Higher gauge theory I: 2-bundles, PHD Thesis, UCR, {\tt math.CT/0410328} (2004).
\bibitem{bell1986} J.L. Bell, From absolute to local mathematics, {\it Synthese}, {\bf 69}, 409-426 (1986).
\bibitem{bell1988} J.L. Bell, {\it Toposes and Local Set Theories,} Clarendon Press, Oxford (1988).
\bibitem{benioff1976} P. Benioff, ''Models of ZF set theory as carriers for the mathematics of physics I, II,'' {\it J. Math. Phys.}, {\bf 19}, 618, 629 (1976).
\bibitem{brans1994} C.H. Brans, {\it J. Math. Phys.} 35, 5494 (1994).
\bibitem{cattaneo2002} A.S. Cattaneo, C.A. Rossi, Wilson surfaces and higher dimensional knot invariants, {\it Commun.Math.Phys.} {\bf 256} (2005) 513-537, {\tt math-ph/0210037}.
\bibitem{crane2006} L. Crane, Categorical Geometry and the Mathematical Foundations of Quantum General Relativity; Contribution to the Cambridge University Press volume on Quantum Gravity, D. Oriti ed. (2006) {\tt gr-qc/0602120}.
\bibitem{feferman1985} S. Feferman, A theory of variable types, {\it Proc. Fifth Latin American Symp. on Math. Logic} eds. X. Caicedo, N.C.A. da Costa, R. Chuaqui, 1985.
\bibitem{degroote2005} H.S. de Groote, Observables I: Stone Spectra, {\tt math-ph/0509020} (2005). 
\bibitem{isham2007a} A. D{\"o}ring and C.J. Isham. A topos foundations for theories of physics: I. Formal languages for physics, {\tt quant-ph/0703060} (2007).
\bibitem{isham2007b} A. D{\"o}ring and C.J. Isham. A topos foundations for theories of physics: II. Daseinissation and liberation of quantum theory, {\tt quant-ph/0703062} (2007).
\bibitem{isham2007c} A. D{\"o}ring and C.J. Isham. A topos foundations for theories of physics: III. The representation of physical quantities with arrows $\delta^{0}(A):\underline{\Sigma}\rightarrow \mathbb{R}^{\geq}$, {\tt quant-ph/0703064} (2007).
\bibitem{isham2007d} A. D{\"o}ring and C.J. Isham. A topos foundations for theories of physics: IV. Categories of systems, {\tt quant-ph/0703066} (2007).
\bibitem{krol2005} J. Kr{\'o}l, Model theory and the AdS/CFT correspondence, (2005) {\tt hep-th/0506003}.
\bibitem{isham2000a} J.Hamilton, J.Butterfield and C.J.Isham, A topos perspective on the Kochen-Specker theorem: III. Von Neuman algebras as the base category, {\it Int. J. Theor. Phys}, {\bf 39}, 1413-1436 (2000).
\bibitem{jackson2006} M. Jackson, A sheaf theoretic approach to measure theory. PhD Thesis,
University of Pittsburgh (2006).
\bibitem{kapustin2000} A. Kapustin, D-branes in topologically nontrivial B field, {\it Adv. Theor. Math. Phys.} {\bf 4} (2000).
127-154. hep-th/9909089.
\bibitem{krol2006a} J. Kr\'ol, A model for spacetime. The role of interpretations in some Grothendieck topoi, {\it Found. Phys.} {\bf 36}(7) (2006).
\bibitem{krol2006} J. Kr\'ol, A model for spacetime II. The emergence of higher dimensions and string/field theory dualities, {\it Found. Phys.} {\bf 36}(7) (2006).
\bibitem{krol2007} J. Kr\'ol, Topos theory and spacetime structure, {\it Int. J. Geom. Meth. Modern Physics} {\bf 4} (3) (2007).
\bibitem{krol2008a} J. Kr\'ol, 2-groups, gerbes and the fundamental symmetry of gravity, in preparation (2008).
\bibitem{lambek1994} J. Lambek and P.J. Scott, Introduction to higher order categorical logic, Cambridge U. Press, 1994.
\bibitem{padmanabhan2008}, T. Padmanabhan, Emergent gravity and dark energy, {\tt gr-qc/0802.1798} (2008).
\item A. Kock, Synthethic Differential Geometry., London Math. Sci. Lecture. Notes Series. 51, Cambridge U. Press 1981. 
\bibitem{maclane1992} S. Mac Lane, I. Moerdijk, {\it Sheaves in Geometry and Logic. A First Introduction to Topos Theory,} (Springer, New York 1992).
\bibitem{moerdijk1991} I. Moerdijk and G. E. Reyes, Models for Smooth infinitesimal Analysis., Springer Verlag N. Y., 1991.
\bibitem{sladkowski2001} J. S{\l}adkowski, Gravity on exotic $R^4$'s with few symmetries, {\it Int. J. Mod. Phys.} {\bf D10}, 311 (2001).
\bibitem{sladkowski1999} J. S{\l}adkowski, {\it Acta Phys. Polon.} B30, 3485 (1999).
\bibitem{straumann2002} N. Straumann, On the cosmological constant problems and the astronomical evidence for a homogeneous energy density with negative pressure, {\tt astro-ph/0203330} (2002).
\bibitem{takeuti1978} G. Takeuti, Two applications of logic to mathematics, {\it Math. Soc. Japan} {\bf 13}, Kano Memorial Lec. 3 (1978).
\bibitem{takeuti1983} G. Takeuti, Quantum logic and quantization, {\it Proc. Int. Symp. Foundations of Quantum Mechanics,} pp. 256-260, (Tokyo, 1983).
\bibitem{weinberg1989} S. Weinberg, {\it Rev. Mod. Phys.} {\bf 61}, 1 (1989).

\end{list}

\end{document}